\documentclass[a4paper,twocolumn,superbib,nofootinbib,showkeys,superscriptaddress,amsmath,amssymb]{revtex4}

\usepackage{color,graphicx}
\usepackage{bm}
\usepackage{url}

\definecolor{dgreen}{rgb}{0,.6,0}
\begin{document}

\title{Breaking a chaos-noise-based secure communication scheme}
\thanks{This paper has been published in \textit{Chaos} (an AIP journal),
vol. 15, no. 1, article 013703 (10 pages), March 2005.}

\author{Shujun Li}
\email[E-mail address: ]{hooklee@mail.com}%
\homepage[personal web site: ]{http://www.hooklee.com.}%
\affiliation{Department of Electronic Engineering, City University
of Hong Kong, Kowloon, Hong Kong SAR, China}
\author{Gonzalo \'{A}lvarez}
%\email[E-mail address: ]{gonzalo@iec.csic.es}%
\affiliation{Instituto de F\'{\i}sica Aplicada, Consejo Superior
de Investigaciones Cient\'{\i}ficas, Serrano 144---28006 Madrid,
Spain}
\author{Guanrong Chen}
%\email[E-mail address: ]{eegchen@cityu.edu.hk}%
\affiliation{Department of Electronic Engineering, City University
of Hong Kong, Kowloon, Hong Kong SAR, China}
\author{Xuanqin Mou}
%\email[E-mail address: ]{xqmou@xjtu.edu.cn}%
\affiliation{School of Electronics and Information Engineering,
Xi'an Jiaotong University, Xi'an, Shaanxi 710049, China}

\begin{abstract}
This paper studies the security of a secure communication scheme
based on two discrete-time intermittently-chaotic systems
synchronized via a common random driving signal. Some security
defects of the scheme are revealed: 1) the key space can be
remarkably reduced; 2) the decryption is insensitive to the
mismatch of the secret key; 3) the key-generation process is
insecure against known/chosen-plaintext attacks. The first two
defects mean that the scheme is not secure enough against
brute-force attacks, and the third one means that an attacker can
easily break the cryptosystem by approximately estimating the
secret key once he has a chance to access a fragment of the
generated keystream. Yet it remains to be clarified if
intermittent chaos could be used for designing secure chaotic
cryptosystems.
\end{abstract}

\keywords{secure communication, encryption, cryptanalysis,
intermittent chaos, noise, synchronization}

\maketitle

\bfseries

In \cite{Minai:ChaosNoiseCipher:Chaos1998}, a new method of
generating pseudo-random keys from discrete-time chaotic systems
using a noise signal was proposed for designing secure
communication schemes. In this method, the noise signal serves as
a common driving signal to synchronize two dynamical systems and
forces them to be intermittently chaotic, ensuring that they
generate the same pseudo-random keys for both encryption and
decryption. This paper studies the security of the above-referred
secure communication scheme and points out its severe security
defects, showing that the scheme is insecure against both
inexpensive brute-force attacks and known/chosen-plaintext
attacks. At present, it is not yet clear whether or not the
existence of intermittent periodicity is always an essential
defect of this kind of secure chaotic cryptosystems.

\mdseries

\section{Introduction}

The idea of using chaos to design analog secure communication
systems and digital ciphers has provoked a great deal of research
efforts since the early 1990s \cite{Yang:Survey2004,
Alvarez:Survey99, ShujunLi:Dissertation2003}. Meanwhile, security
analysis of various proposed chaotic cryptosystems also attracts
increasing attention, and some chaotic cryptosystems have been
found insecure \cite{Biham:Cryptanalysis:EuroCrypt91,
Beth:ChaoticCryptanalysis:EuroCrypt94,
Perez:ReturnMapCryptanalysis:PRL95, Yang:STZCR:IJCTA95,
Kocarev:BreakingParameters:IJBC96, Kocarev:UnmaskingChaos:IJBC97,
Short:ChaoticSignalExtraction:IJBC97,
Zhou:ExtractChaoticSignal:PLA97, Yang:GSCryptanalysis:IEEETCASI97,
Yang:SpectralCryptanalysis:PLA98, Short:UnmaskingHyperchaos:PRE98,
Ogorzatek:AttackChaoticCiphers:ISCAS98,
ZhouLai:ChaoticCryptanalysis:PRE99a,
ZhouLai:ChaoticCryptanalysis:PRE99b,
Alvarez:ChaoticCryptanalysis:PLA2000,
LiShujun:ChaoticCipher:PLA2001,
Sobhy:AttackingChaoticCipher:ICASSP2001,
Short:ChaoticCrptanalysis:IEEETCASI2001,
Huang:UnmaskingChaosWavlet:IJBC2001,
Alvarez:BreakingBapistaCipher:PLA2003,
Alvarez:BreakingPareeketal:PLA2003,
LiShujun:ChaoticCryptanalysis:IJBC2003,
LiShujun:ChaoticCipher:PLA2003, LiShujun:CPC2003,
TaoDu:ChaoticCrytpanalysis:IJBC2003a, Solak:BreakingHenon:PLA2004,
Alvarez:BreakingHenon:CSF2004, Chee:ZCCryptanalysis:CSF2004,
Wu:ChuaParameterEstimation:PLA2004, Alvarez:BreakingPS:Chaos2004,
Alvarez:BreakingWongCipher:PLA2004,
Alvarez:ChaoticCryptanalysis:JSV2004}. Most analog chaos-based
secure communication systems are based on chaos synchronization
technique \cite{Pecora:Synchronzation90}, where the receiver
(response or slave) chaotic system synchronizes with the
transmitter (drive or master) system via a signal transmitted over
a public channel. After the chaos synchronization is achieved, the
plain-signal can be recovered in different ways corresponding to
different structures of the encryption algorithms. According to
the method used for encrypting the plain-signal, there are four
common types of chaos synchronization-based encryption structures
\cite{Feldmann:ISA:IJCTA96, Alvarez:Survey99, Yang:Survey2004}:
chaotic masking, chaotic switching (or chaotic shift keying -
CSK), chaotic modulation and inverse system approach. Since many
early chaos-based secure communication systems are found insecure
against various attacking methods, some countermeasures have been
proposed in the hope of enhancing the security: 1) using more
complex dynamical systems, such as hyperchaotic systems or
multiple cascaded heterogeneous chaotic systems
\cite{Murali:HeterogeneousChaosCryptography:PLA2002}; 2) adding
traditional ciphers into the whole cryptosystem
\cite{Yang-Wu-Chua:ChaoticCryptography:IEEETCASI1997}; 3)
introducing a discrete-time impulsive signal instead of a
continuous signal to realize synchronization
\cite{Yang-Chua:ImpulsiveChaos:IJBC1997, Yang:Survey2004}. The
first countermeasure has been found not secure enough against some
attacks \cite{Short:UnmaskingHyperchaos:PRE98,
ZhouLai:ChaoticCryptanalysis:PRE99b,
Huang:UnmaskingChaosWavlet:IJBC2001,
TaoDu:ChaoticCrytpanalysis:IJBC2003a}, and some security defects
of the second have also been reported
\cite{Short:ChaoticCrptanalysis:IEEETCASI2001}, but the last one
has not yet been broken to date.

In \cite{Minai:NoiseSynchronization:PRE1998}, as a new way to
enhance the security of chaos-based secure communications, a new
synchronization method was proposed for multiple discrete-time
dynamical systems driven by a common random signal. Dynamical
systems synchronized in this way work in intermittent chaotic
states, i.e., alternatively switching between chaotic and periodic
regimes. In \cite{Minai:ChaosNoiseCipher:Chaos1998}, this new
synchronization method was used to construct a new secure
communication scheme, in which the two synchronized
intermittently-chaotic systems generate the same pseudo-random
keystream. The pseudo-random keystream is then used for encrypting
the plaintext at the transmitter end, and for decrypting the
ciphertext at the receiver, via a piecewise linear
encryption/decryption function originally used in
\cite{Yang-Wu-Chua:ChaoticCryptography:IEEETCASI1997}. The core of
this secure communication scheme is the key-generation process
based on the two discrete-time intermittently-chaotic systems. In
\cite{Minai:ChaosNoiseCipher:Chaos1998}, it was claimed that the
key-generation process is very sensitive to the mismatch of the
secret key so it is very secure against attacks. To the best of
our knowledge, this secure communication scheme is the first and
the only one based on intermittent chaos, and it has not been
cryptanalyzed before.

This paper analyzes the security of the above-referred secure
communication scheme, and reveals some of its security defects: 1)
some secret parameters (i.e., sub-keys) can be eliminated or
directly estimated from the ciphertext, so that the available
key-space is drastically reduced; 2) the decryption is largely
insensitive to the mismatch of the secret key; 3) the chaos-based
key-generation process is insecure against known/chosen-plaintext
attacks. The first two defects mean that this communication scheme
is not secure enough against brute-force attacks, and the third
defect means that an attacker can easily break the cryptosystem by
approximately estimating the secret key once he has a chance to
get access to a fragment of the generated keystream. It is not
clear, at this stage, whether or not intermittent chaos is always
unsuitable for designing secure chaotic cryptosystems.

The rest of this paper is organized as follows. In the next
section, a brief introduction to the secure communication scheme
under study is given. In Sec. \ref{section:Cryptanalysis},
cryptanalysis is shown in detail along with experimental
verification. The last section concludes the paper.

\section{The Secure Communication Scheme}

The secure communication system proposed in
\cite{Minai:ChaosNoiseCipher:Chaos1998} is described as follows.

\textit{The transmitter system} is
\begin{equation}
z_{t+1}^{(1)}=\tanh(\mu(az_t^{(1)}+u_t))-\tanh(\mu bz_t^{(1)}),
\label{equation:TransmitterSystem}
\end{equation}
and \textit{the receiver system} is
\begin{equation}
z_{t+1}^{(2)}=\tanh(\mu(az_t^{(2)}+u_t))-\tanh(\mu
bz_t^{(2)}),\label{equation:ReceiverSystem}
\end{equation}
where $\mu,a,b$ are positive parameters. The synchronization
between these two discrete-time dynamical systems is realized via
a common driving signal,
\begin{equation}
u_t=\phi(\hat{r}_t)=\phi(r_t+\mu_t)=\begin{cases} A, & \hat{r}_t<\theta,\\
B, & \hat{r}_t\geq \theta,\\
\end{cases}
\end{equation}
where $r_t$ is a random telegraph signal (RTS) given by
\begin{equation} r_t=\begin{cases}\alpha, & \mbox{with
probability }p,\\\beta, & \mbox{with probability }1-p,
\end{cases}
\end{equation}
and $\mu_t$ is noise introduced by the channel during the
transmission of $r_t$ (note that $\mu_t$ may be different at the
sender and the receiver ends). To facilitate the following
descriptions, without loss of generality, assume $0<\alpha<\beta$
and $0<A<B$. Then, a typical value of $\theta$ is
$(\alpha+\beta)/2$, which is the middle value of the interval
$[\alpha,\beta]$\footnote{Note that Eq. (9) in
\cite{Minai:ChaosNoiseCipher:Chaos1998},
$\theta=\alpha+(\alpha+\beta)/2$, is wrong, since
$\alpha+(\alpha+\beta)/2>\beta$ when $\alpha<\beta<2\alpha$. In
fact, under the assumption that noise $\mu_t$ has zero-mean and
symmetric distribution \cite{Minai:ChaosNoiseCipher:Chaos1998},
the best value of $\theta$ should naturally be the middle value of
$[\alpha,\beta]$.}.

In the above secure communication scheme, the RTS signal $r_t$
randomly switches the dynamics of the two discrete-time maps
(\ref{equation:TransmitterSystem}) and
(\ref{equation:ReceiverSystem}) by changing the value of one
control parameter, $u_t$. According to
\cite[Sec.II]{Minai:ChaosNoiseCipher:Chaos1998}, when $\mu$, $a$
and $b$ are set to make the two maps chaotic for $u_t=0$,
increasing $u_t$ will cause the maps to go into the periodic
regime. Thus, by choosing the values of $A$ and $B$ properly, the
two synchronized systems can be configured to work in the chaotic
regime for $u_t=A$ and in the periodic regime for $u_t=B$,
respectively. That is, the two maps are not fully chaotic, but
\textit{intermittently chaotic}, under the control of the random
signal $u_t$. In this case, the two systems can reach
synchronization after a limited number of iterations if $p$ is not
too large \cite{Minai:NoiseSynchronization:PRE1998}. According to
our experiments, $p\leq 0.8$ is required when
$\mu=5,a=1,b=1,A=\alpha=0.02,B=\beta=0.2$ (the default values used
in \cite{Minai:ChaosNoiseCipher:Chaos1998}).

\textit{The encryption procedure} is composed of two processes:
the key-generation process and the encryption process. Similarly,
\textit{the decryption procedure} is composed of the
key-generation process and the decryption process. Both procedures
can be described as follows.

\begin{itemize}
\item \textit{The key-generation process}: the transmitter and the
receiver generate two pseudo-random keystreams,
$\{K_t^1=z_t^{(1)}\}$ and $\{K_t^2=z_t^{(2)}\}$, respectively.

\item Assuming the plain-signal is $m_t$ and the transmitted
cipher-signal is $s_t$, \textit{the encryption procedure} is
$s_t=\Psi^n\left(m_t,K_t^1\right)$, where $\Psi(x,y)$ is a
piecewise linear encryption function originally used in
\cite{Yang-Wu-Chua:ChaoticCryptography:IEEETCASI1997}:
\begin{equation}
\Psi(x,y)=\begin{cases}
(x+y)+2w, & -3w\leq x+y\leq -w,\\
(x+y), & -w\leq x+y\leq w,\\
(x+y)-2w, & w\leq x+y\leq 3w.
\end{cases}
\label{equation:nShiftCipher}
\end{equation}
As mentioned in \cite{Minai:ChaosNoiseCipher:Chaos1998},
$\Psi^n(x,y)$ can be replaced by other encryption functions.

\item Assuming the received cipher-signal is
$\hat{s}_t=s_t+\delta_t$, where $\delta_t$ is the noise introduced
in the transmission channel, and the recovered plain-signal is
$\hat{m}_t$, \textit{the decryption procedure} is
$\hat{m}_t=\Psi^n\left(\hat{s}_t,-K_t^2\right)$.
\end{itemize}

\textit{The secret key} of the key-generation process is
$\{\mu,a,b,A,B,\theta\}$ and \textit{the secret key} of the
encryption function (\ref{equation:nShiftCipher}) is $\{n,w\}$.

\textit{An enhanced key-generation process} was also proposed to
improve the security of the generated key-stream: use $k\;(k>1)$
different systems instead of a single one at both ends, and take
the maximal value of the outputs of all $k$ systems at each time
instant to determine $K_t^1$ and $K_t^2$.

\section{Cryptanalysis}
\label{section:Cryptanalysis}

Before starting to analyze the security of the above-referred
secure communication scheme, some security guidelines are
reviewed. Following the well-known Kerckhoffs' principle in
cryptology \cite{Schneier:AppliedCryptography96}, the security of
a cryptosystem should rely on the secret key only, which means
that an attacker knows all details about the cryptosystem except
for the secret key. For this secure communication scheme to be
analyzed, the following assumptions are made:
\begin{itemize}
\item the attacker does not know the secret keys
$\{\mu,a,b,A,B,\theta\}$ and $\{n,w\}$.

\item the attacker exactly knows Eqs.
(\ref{equation:TransmitterSystem}) to
(\ref{equation:nShiftCipher});

\item the attacker has full control on the channel over which the
cipher-signals $s_t,\hat{s}_t$ and the common driving signals
$r_t,\hat{r}_t$ are transmitted, where ``full control" means that
the attacker not only can passively observe the transmitted
signals but also can actively change the transmitted signals $s_t$
and $r_t$.
\end{itemize}
Actually, in some special scenarios, it is possible for an
attacker to get some useful information or even intentionally
choose some information from the transmitter and/or the receiver.
As a result, from the cryptographical point of view, to provide a
high level of security, a cryptosystem should be secure enough
against all the following four attacks (listed from the hardest to
the easiest):
\begin{itemize}
\item \textit{the ciphertext-only attack} - the attacker can only
get ciphertexts and other publicly-transmitted information (such
as the common driving signal in the scheme under discussion);

\item \textit{the known-plaintext attack} - in addition to some
basic information, the attacker can get some plaintexts and the
corresponding ciphertexts;

\item \textit{the chosen-plaintext attack} - in addition to some
basic information, the attacker can choose some plaintexts and get
the corresponding ciphertexts;

\item \textit{the chosen-ciphertext attack} - in addition to some
basic information, the attacker can choose some ciphertexts and
get the corresponding plaintexts.
\end{itemize}
The last two attacks, which seem to seldom occur in practice, are
feasible in some real applications \cite[Sec.
1.1]{Schneier:AppliedCryptography96} and has become much more
common in today's networked world. In the following, it will be
pointed out that the secure communication system under study is
not secure enough against the first three attacks.

This paper imposes a simple assumption, $\mu_t\equiv 0$ and
$\delta_t\equiv 0$, to make the discussion of cryptanalysis
easier. Note that removing the two noise signals has no influence
on the security analysis of the studied scheme. Also, $\mu_t$ and
$\delta_t$ can be directly removed in some applications, for
example, when the whole system is constructed digitally in
computers and the transmission channels are digital storage media
(such as floppy disks, hard disks, CDs, flash-memory disks, etc.)
or networks completely digitized with some error-correction
mechanisms. In addition, one generally simulates secure
communication schemes via some mathematical softwares, such as
Matlab$^\circledR$, where no channel errors occur.

\subsection{Reduction of the key space}

The simplest attack to a cryptosystem is known as the brute-force
attack consisting of exhaustively searching all possible keys. The
complexity of such a simple attack is determined by the size of
the key space, i.e., the number of all valid keys. From the
cryptographical point of view, the size of the key space should
not be smaller than $2^{100}$ to provide a high level of security
\cite{Schneier:AppliedCryptography96}. In this section, it is to
point out that the studied communication system is not secure
enough since its key space is not sufficiently large, which is
caused by the key space reduction and low sensitivity of
decryption to the secret key.

The secret key of the key-generation process
$\{\mu,a,b,A,B,\theta\}$ can be immediately reduced to
$\{a_\mu=\mu a,b_\mu=\mu b,A_\mu=\mu A,B_\mu=\mu B,\theta\}$ by
rewriting the chaotic equation as follows:
\begin{equation}
z_{t+1}=\tanh\left(a_\mu z_t+u_{\mu,t}\right)-\tanh\left(b_\mu
z_t\right),
\end{equation}
where
\begin{equation}
u_{\mu,t}=\mu\phi(\hat{r}_t)=\begin{cases} \mu A=A_\mu, & \hat{r}_t<\theta,\\
\mu B=B_\mu, & \hat{r}_t\geq \theta.\\
\end{cases}
\end{equation}
In addition, as stated in \cite{Minai:ChaosNoiseCipher:Chaos1998},
this secure communication scheme is not sensitive to $\theta$.
Actually, $\theta$ depends only on the distribution of the noise
signal $\mu_t$ and under most conditions
$\overline{\theta}=(\alpha+\beta)/2$ can work well. Therefore,
from the cryptographical point of view, $\theta$ should not be
included in the secret key. Thus, the secret key of the
key-generation process is reduced to
$\{a_\mu,b_\mu,A_\mu,B_\mu\}$.

The secret key of the encryption function
(\ref{equation:nShiftCipher}) was claimed to be $\{n,w\}$ in
\cite{Minai:ChaosNoiseCipher:Chaos1998}. Since the range of the
cipher-signal $s_t$ is $[-w,w]$, one can set $w$ be the maximum of
$|s_t|$ in a long period of time. This means that $w$ can be
removed and the secret key is further reduced to $\{n\}$ only.

In the following, only $\{a_\mu,b_\mu,A_\mu,B_\mu,n\}$ will be
used as the secret parameters of the secure communication scheme
for further analysis.

\subsection{The ``plausible" sensitivity of the decryption
error to parameter mismatch}

In \cite{Minai:ChaosNoiseCipher:Chaos1998}, some experiments were
reported to show that the secure communication scheme is very
sensitive to parameter mismatch: even a small difference of order
$10^{-8}$ in $\mu$, $\alpha$ or $\beta$ will cause a relatively
large decryption error (see Fig. 6 of
\cite{Minai:ChaosNoiseCipher:Chaos1998}). Assuming such a
sensitivity holds equally for all the four secret parameters
$\{a_\mu,b_\mu,A_\mu,B_\mu\}$, one can see that the size of the
key space of the key-generation process will not be less than
$(10^{8})^4=10^{32}\approx 2^{106}$, which is cryptographically
large. However, our numerical study shows that the data given in
Fig. 6 of \cite{Minai:ChaosNoiseCipher:Chaos1998} are wrong, so
they reached a false conclusion on the sensitivity to the secret
key in decryption. In fact, we found that the sensitivity is
mainly dependent on the values of $p$ and $n$, and that this
sensitivity is too weak to provide a high level of security when
$n$ is not too large ($n$ has to be as large as $2^{32}$ as
discussed below).

\begin{figure}[!htbp]
\includegraphics{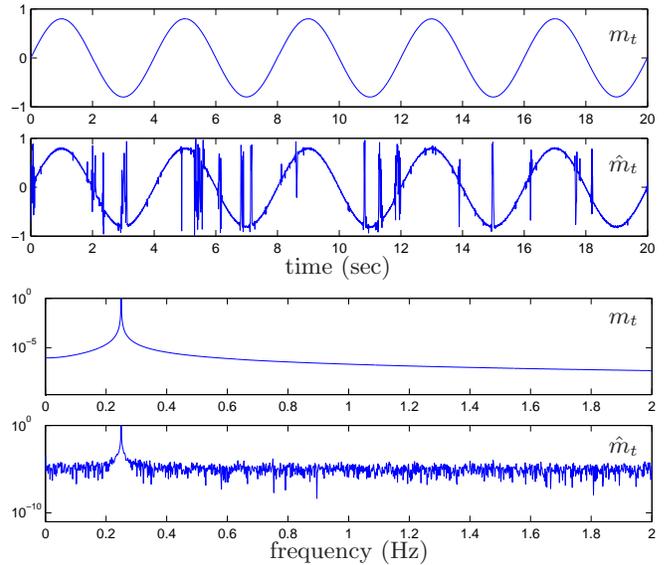}
\caption{The waveforms of the original plain-signal
$m_t=0.8\sin(2\pi t/4)$ and the decrypted signal $\hat{m}_t$, and
their relative power spectra, when $\delta=0.001$, $p=0.3$ and
$n=71$.}\label{figure:mt_mt2_sint_3}
\end{figure}

\begin{figure}[!htbp]
\includegraphics{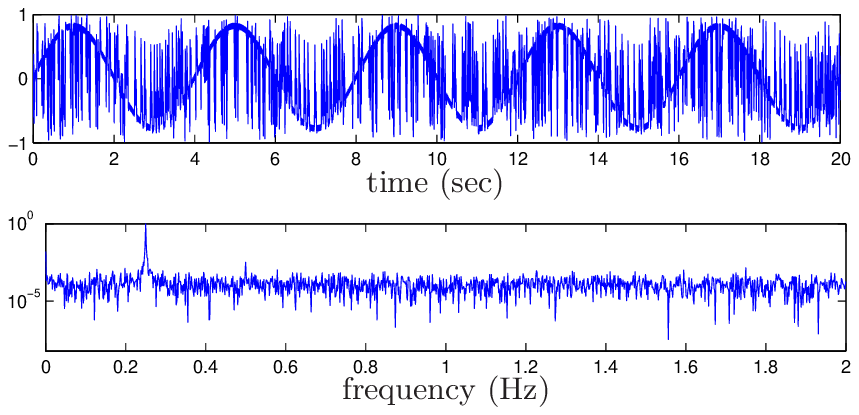}
\caption{The decrypted plain-signal $\hat{m}_t$ and its relative
power spectrum, when $m_t=0.8\sin(2\pi t/4)$, $\delta=1$, $p=0.3$
and $n=71$.}\label{figure:mt_mt2_sint_0}
\end{figure}

\begin{figure}[!htbp]
\includegraphics{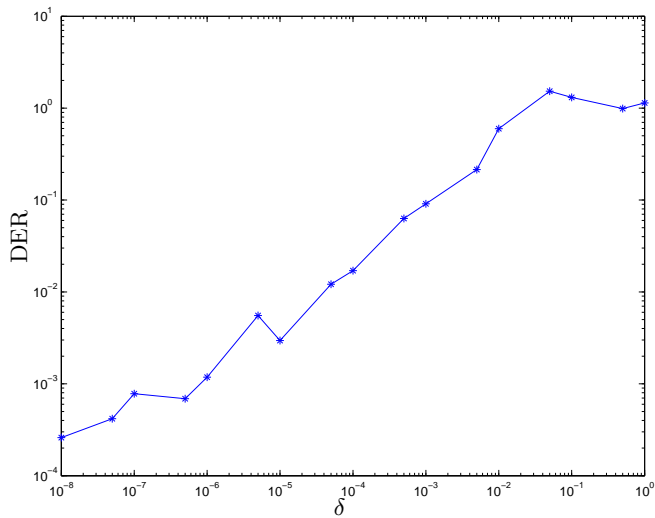}
\caption{\emph{The experimental relationship between DER and the
value of $\delta$, when $p=0.3$ and
$n=71$.}}\label{figure:PER_delta}
\end{figure}

At first, let us revisit the case experimentally studied in Sec.
III-C of \cite{Minai:ChaosNoiseCipher:Chaos1998}, where the
key\footnote{Note that the value of $b$ shown in the caption of
Fig. 5 of \cite{Minai:ChaosNoiseCipher:Chaos1998} should be 1.0,
not 5.0. In fact, the default secret parameters used in
\cite{Minai:ChaosNoiseCipher:Chaos1998} are always $\mu=5,a=5,b=1$
(5/5/1-oscillator).} is
$\{a_\mu=25,b_\mu=5,A_\mu=0.05,B_\mu=1,n=71\}$, $w=1$, and the
plain-signal is $m_t=0.8\sin(2\pi t/4)$. In
\cite{Minai:ChaosNoiseCipher:Chaos1998}, it was not explicitly
mentioned what the sampling frequency is. Here, we assume that the
plain-signal is sampled at a frequency of 100 Hz. When the
decryption key is
$\{a_\mu'=a_\mu(1+\delta),b_\mu'=b_\mu(1+\delta),A_\mu'=A_\mu(1+\delta),B_\mu'=B_\mu(1+\delta)\}$,
where $\delta=0.001$ is the relative parameter
mismatch\footnote{Since $A_\mu$ may be very small in practice, the
relative mismatch is more suitable for analysis than its absolute
counterpart.}, the decrypted plain-signal $\hat{m}_t$ and its
spectrum are shown in Fig. \ref{figure:mt_mt2_sint_3} (for
comparison, the waveform and spectrum of $m_t$ are also plotted).
One can see that the plain-signal is decrypted with only a little
intermittent noise, which means that the sensitivity of the
encryption scheme to the parameter mismatch is very weak. Further
experiments show that with a larger $\delta$ it is still possible
to approximately recover the plain-signal $m_t$, at least in the
frequency domain (see Fig. \ref{figure:mt_mt2_sint_0} for the
decryption plain-signal $\hat{m}_t$ when $\delta=1$). What does
$\delta=1$ mean? It means that even a secret key satisfying
$a'_\mu=2a_\mu$, $b'_\mu=2b_\mu$, $A'_\mu=2A_\mu$ and
$B'_\mu=2B_\mu$ can be used to approximately recover the
sinusoidal signal. To observe the relationship between the
recovery errors and the value of $\delta$, defining the decryption
error ratio (in power energy) as
$\mathrm{DER}=\left(\sum_t(m_t'-m_t)^2\right)/\sum_tm_t^2$, 17
different values of $\delta$ have been tested and one experimental
result\footnote{All experiments show similar results, so only one
is plotted here.} is shown in Fig.~\ref{figure:PER_delta}. The
results imply that any key satisfying $a_\mu/2\leq a'_\mu\leq
2a_\mu$, $b_\mu/2\leq b'_\mu\leq 2b_\mu$, $A_\mu/2\leq A'_\mu\leq
2A_\mu$ and $B_\mu/2\leq B'_\mu\leq 2B_\mu$ can be used to
approximately recover the sinusoidal signal. Roughly speaking, the
closer the key to the real key, the smaller the recovery errors
will incline to be. Clearly, this will cause fatal collapse of the
key space and so dramatically reduce the security of the scheme.
Thus, one question arises: how can one explain such an undesirable
insensitivity?

\begin{figure}[!htbp]
\includegraphics{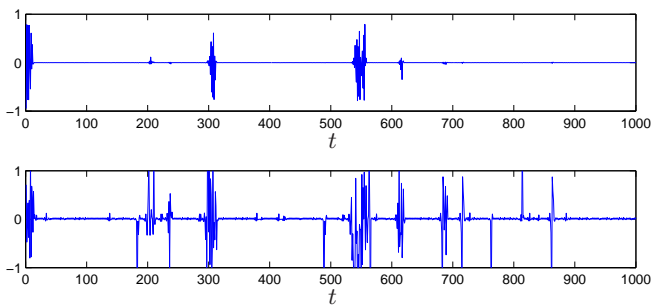}
\caption{The two difference signals, $K_t^2-K_t^1$ (top) and
$\Psi^n(0,K_t^2)-\Psi^n(0,K_t^1)$ (bottom), when $\delta=0.001$,
$p=0.3$ and $n=71$.}\label{figure:Kt12_3}
\end{figure}

In our opinion, the reason behind this interesting phenomenon
should be attributed to the fact that the involved dynamical
systems show chaotic behaviors with a probability less than $p$
\cite{Minai:NoiseSynchronization:PRE1998}. When $p<0.5$, the
periodic behavior will dominate the evolution of $K_t$, therefore
$K_t$ will not be so sensitive to parameter mismatch in an average
sense, as expected in a fully chaotic regime. Observing the
difference between $K_t^2$ and $K_t^1$, shown in Fig.
\ref{figure:Kt12_3}, the above explanation can be understood
conceptually.

\begin{figure}[!htbp]
\includegraphics{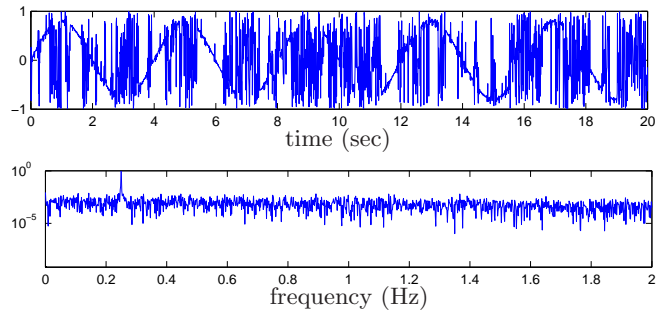}
\caption{The decrypted plain-signal $\hat{m}_t$ and its relative
power spectrum, when $m_t=0.8\sin(2\pi t/4)$, $\delta=0.001$,
$p=0.7$ and $n=71$.}\label{figure:mt_mt2_sint_3_p7}
\end{figure}

\begin{figure}[!htbp]
\includegraphics{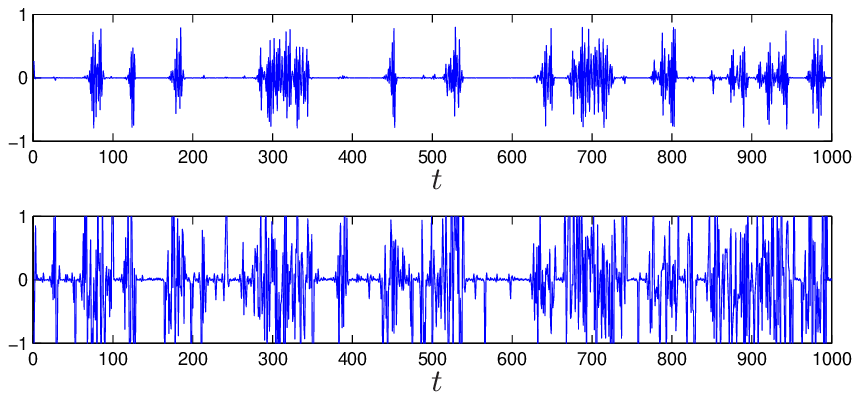}
\caption{The two difference signals, $K_t^2-K_t^1$ (top) and
$\Psi^n(0,K_t^2)-\Psi^n(0,K_t^1)$ (bottom), when $\delta=0.001$,
$p=0.7$ and $n=71$.}\label{figure:Kt12_3_p7}
\end{figure}

Next, let us find out how the sensitivity changes as $p$
increases. Following the above qualitative explanation on the low
sensitivity of the decryption error for $p=0.3$, it can be
expected that the chaotic behavior occurs more and more frequently
as $p$ increases, so that the decryption error will become more
and more sensitive to parameter mismatch. Taking $p=0.7$ as an
example, experiments show that the decryption error when
$\delta=0.001$ is similar to the case when $p=0.3$ and $\delta=1$
(see Figs. \ref{figure:mt_mt2_sint_3_p7} and
\ref{figure:Kt12_3_p7} for the experimental results and compare
them with Figs. \ref{figure:mt_mt2_sint_3} and
\ref{figure:Kt12_3}). Further experiments have been carried out to
check on some other values of $p$, and the results are summarized
in Table \ref{table:delta_max_p}. Note that the synchronization
will become too slow or even impossible when $p>0.8$
\cite{Minai:NoiseSynchronization:PRE1998}. From the data listed in
Table \ref{table:delta_max_p}, it is clear that even for the
maximal value of $p$, i.e., $p=0.8$, the sensitivity of the
decryption error to parameter mismatch is not sufficiently high.
To avoid such an insensitivity, one has to ensure the dynamical
system evolves in the chaotic regime in all time, i.e., to set
$p=1$. However, in this case the synchronization will become
absolutely impossible \cite{Minai:NoiseSynchronization:PRE1998}.
This seems to imply that the proposed secure communication scheme
based on intermittent chaos is \textit{always} insecure from the
cryptographical point of view
\cite{Schneier:AppliedCryptography96}. Yet, it remains to
theoretically clarify whether or not the above claim is right for
all cryptosystems based on intermittently chaotic systems.

\begin{table}[!htbp]
\caption{The largest insensitive parameter mismatch
$\delta_{\max}$ with respect to $p$.}\label{table:delta_max_p}
\begin{ruledtabular}
\begin{tabular}{*{9}{c|}c}
$p$ & 0.1 & 0.2 & 0.3 & 0.4 & 0.5 & 0.6 & 0.7 & 0.8 & $>0.8$\\
\hline $\delta_{\max}$ & 1 & 1 & 1 & $10^{-2}$ & $10^{-2}$ &
$10^{-3}$ & $10^{-3}$ & $10^{-4}$ & impractical
\end{tabular}
\end{ruledtabular}
\end{table}

\begin{figure}[!htbp]
\includegraphics{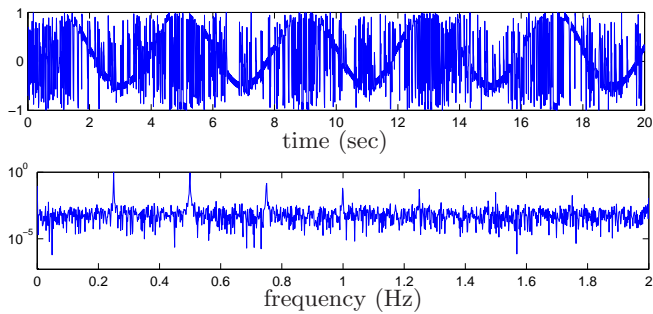}
\caption{The decrypted plain-signal $\hat{m}_t$ and its relative
power spectrum, when $m_t=0.8\sin(2\pi t/4)$, $\delta=10^{-8}$,
$p=0.3$ and $n=2^{31}$.}\label{figure:mt_mt2_sint_3_n31}
\end{figure}

\begin{figure}[!htbp]
\includegraphics{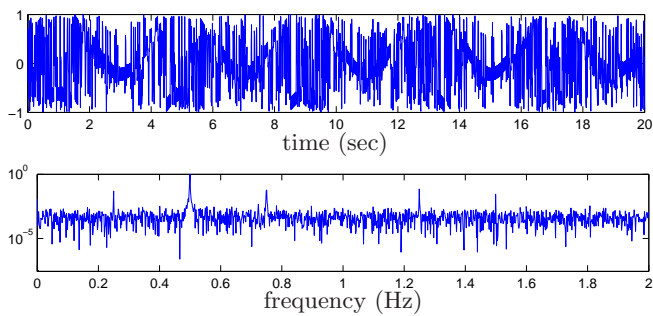}
\caption{The decrypted plain-signal $\hat{m}_t$ and its relative
power spectrum, when $m_t=0.8\sin(2\pi t/4)$, $\delta=10^{-8}$,
$p=0.3$ and $n=2^{32}$.}\label{figure:mt_mt2_sint_3_n32}
\end{figure}

Another fact one can find from Figs.~\ref{figure:Kt12_3} and
\ref{figure:Kt12_3_p7} is that the difference between $K_t^2$ and
$K_t^1$ is significantly magnified by the encryption function
(\ref{equation:nShiftCipher}). By rewriting the $n$-fold
encryption function as $\Psi^n(x,y)=\Psi'(x+ny)$, where
\begin{equation*}
\Psi'(x)=\begin{cases}
((x+w)\bmod 2w)-w, & \left((x/w)\bmod 4\right)\neq 3,\\
w, & \left((x/w)\bmod 4\right)=3,
\end{cases}
\end{equation*}
one can easily find that such a magnification is mainly determined
by the multiplication factor $n$: the larger the $n$ is, the
larger the magnification will be. This suggests that the
sensitivity of $K_t$ is improved by using a larger value of $n$.
However, $n$ has to be very large to increase the sensitivity to
an acceptable level of security since the magnification spreading
rate here is linear. For example, when $p=0.3$, to make the actual
sensitivity be in the order of $10^{-8}$, as reported in
\cite{Minai:ChaosNoiseCipher:Chaos1998}, $n>2^{32}$ is required.
Figs.~\ref{figure:mt_mt2_sint_3_n31} and
\ref{figure:mt_mt2_sint_3_n32} show the decryption results when
$n=2^{31}$ and $n=2^{32}$, respectively. The sinusoid signal $m_t$
can still be distinguished when $n=2^{31}$ whereas the spectral
peak of $m_t$ at 0.25 Hz is suppressed when $n=2^{32}$ (but the
signal is still partially visible). This result actually implies:
\begin{itemize}
\item The decryption is largely insensitive not only to the
mismatch of the secret key of the key-generation process, but also
to the secret key of the encryption function
(\ref{equation:nShiftCipher}), so brute-force attacks to the whole
chaotic cryptosystem are quite easy.

\item \textit{The security of the studied communication scheme is
ensured by the encryption function $\Psi^n(x,y)$, not by the
chaos-based key-generation process}. In other words, if the
key-generation process is directly used as a keystream generator
to encrypt the plain-signal, the cryptosystem will be rather weak.
This means that the chaos-based key-generation process is
irrelevant to the security of the studied secure communication
scheme.
\end{itemize}

Finally, it should be mentioned that the experimental data given
in \cite{Minai:ChaosNoiseCipher:Chaos1998} are actually wrong.
When $p=0.3,n=71$, our experiments show that the power energy of
the decryption error is much smaller than the one shown in Fig. 6
of \cite{Minai:ChaosNoiseCipher:Chaos1998}. For example, when
$\delta=0.001$, the power energy is only 0.093086. It is found
that the data for $p=1-0.3=0.7$ agree with Fig. 6 of
\cite{Minai:ChaosNoiseCipher:Chaos1998}. We guess that the authors
of \cite{Minai:ChaosNoiseCipher:Chaos1998} mistook the value of
$p$.

\subsection{Breaking the key-generation process by known/chosen-plaintext attacks}

\begin{figure}[!htbp]
\includegraphics{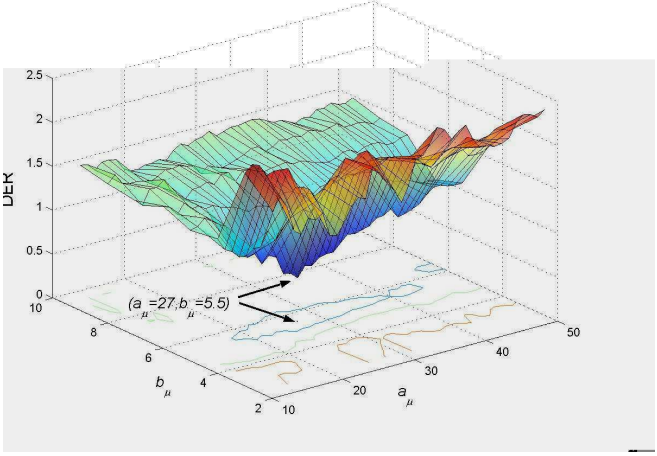}
\caption{DER vs. $\{a_\mu,b_\mu\}$, when
$A_\mu=0.06,B_\mu=1$.}\label{figure:DER1}
\end{figure}

\begin{figure}[!htbp]
\includegraphics{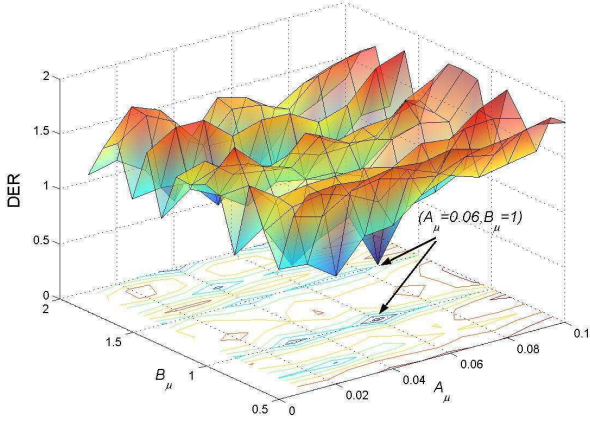}
\caption{DER vs. $\{A_\mu,B_\mu\}$, when
$a_\mu=27,b_\mu=5.5$.}\label{figure:DER2}
\end{figure}

\begin{figure}[!htbp]
\includegraphics{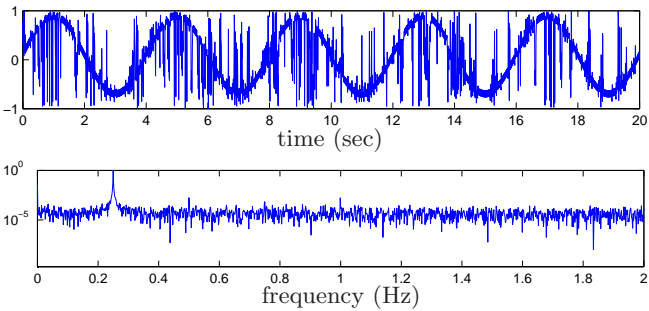}
\caption{The decrypted plain-signal $\hat{m}_t$ and its relative
power spectrum with the estimated parameters
$\{\widetilde{a}_\mu=27,\widetilde{b}_\mu=5.5,\widetilde{A}_\mu=0.06,\widetilde{B}_\mu=1\}$
in a real attack, when $m_t=0.8\sin(2\pi t/4)$, $\delta=10^{-8}$,
$p=0.3$ and $n=71$.}\label{figure:mt_mt2_sint_3_attack}
\end{figure}

In a known/chosen-plaintext attack, the attacker can get the
plain-signal $m_t$, so it is possible to find an approximate key
$\{a_\mu,b_\mu,A_\mu,B_\mu\}$ by searching the whole key space.
Here, we assume that the value of $n$ is known. As discussed
above, the decryption error is not sensitive to the mismatch of
$\{a_\mu,b_\mu,A_\mu,B_\mu\}$, so the searching complexity will be
small practically. Assuming that the range of the four secret
parameters are
$a_\mu\in[12,50],b_\mu\in[2.5,9.5],A_\mu\in[0.02,0.1],B_\mu\in[0.5,2]$,
respectively, the searching steps are chosen as
$\delta_{a_\mu}=\delta_{b_\mu}=1,\delta_{A_\mu}=0.01,\delta_{B_\mu}=0.1$,
respectively\footnote{In \cite{Minai:ChaosNoiseCipher:Chaos1998},
these ranges are not explicitly given, so we just choose typical
ranges that can ensure the intermittent chaoticity of the sender
and receiver maps.}. Note that the range and the searching step of
$b_\mu$ are intentionally chosen to make sure that the real value
$b_\mu=5$ cannot be visited in the current searching precision,
which is common in real attacks since the real values of the
secret parameters are all unknown. In this case, the number of all
searched keys is $44928\approx 2^{15.4}$, which is very small even
for a PC. For each guessed key, the decryption error ratio in
power energy (DER, see the previous subsection) is calculated.
Using Matlab$^\circledR$ 6.1, about 4.37 hours is consumed on a PC
with a 1.8GHz Pentium$^\circledR$ 4 CPU and 256MB memory to test
all 44,928 keys. The minimal DER occurs when the key is
$\{\widetilde{a}_\mu=27,\widetilde{b}_\mu=5.5,\widetilde{A}_\mu=0.06,\widetilde{B}_\mu=1\}$.
The DER with respect to $\{a_\mu,b_\mu\}$ and $\{A_\mu,B_\mu\}$
are shown in Figs. \ref{figure:DER1} and \ref{figure:DER2},
respectively, from which one can see that the minimum is
sufficiently distinguishable from other values in the current
searching precision. Compared with the real key
$\{a_\mu=25,b_\mu=5,A_\mu=0.05,B_\mu=1\}$, such a key is good
enough to get an acceptable decryption performance (see Fig.
\ref{figure:mt_mt2_sint_3_attack}). Of course, by doing more
rounds of searching in smaller ranges with smaller steps, it is
easy to get a more accurate estimation of the secret key.
Moreover, since the DER function can be continuous, it may be
possible to use some local minimization scheme to hasten the
search further.

\begin{figure*}[!htbp]
\includegraphics{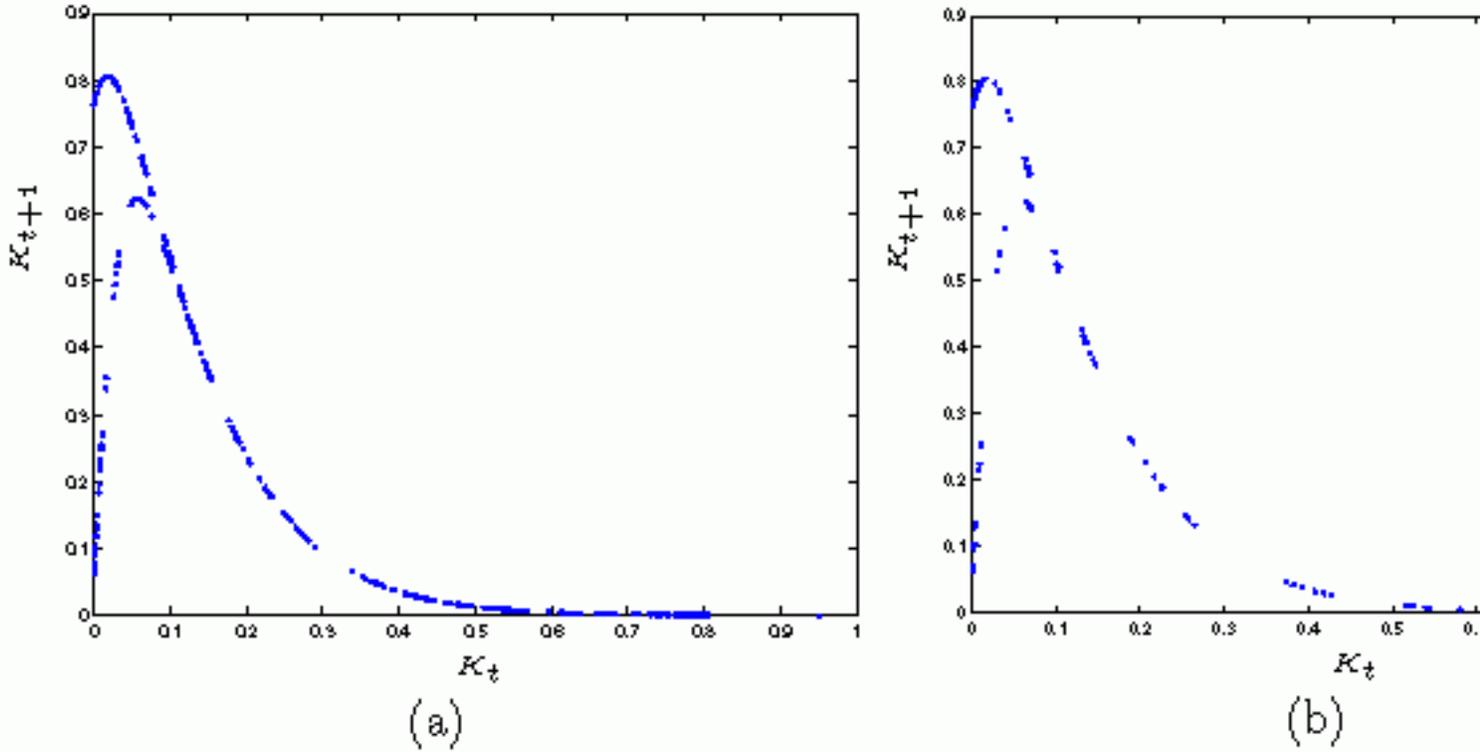}
\caption{The return maps plotted with a variable number of known
samples of $K_t$: (a) 9,000 samples; (b) 1,000 samples; (c) 200
samples.}\label{figure:ReturnMap}
\end{figure*}

In the following, we revisit the security problem studied in Sec.
III.C of \cite{Minai:ChaosNoiseCipher:Chaos1998}: ``\textit{how
secure is the key generator if the intruder has access to a key
fragment, $K_t$, and the corresponding synchronizing signal,
$R_t$, $t_1\leq t\leq t_2$?}" When the $n$-fold encryption
function $\Psi^n(x,y)$ is replaced by another encryption function
that is \textit{invertible with respect to $K_t$}, such as the XOR
operation widely used in cryptography
\cite{Schneier:AppliedCryptography96}, the above attacking
scenario will be possible in known/chosen-plaintext attacks. In
this case, it is obvious that one can immediately construct a
return map by plotting the relationship between $K_{t+1}$ and
$K_t$. When $\mu=5,a=5,b=1,A=0.01,B=0.2$ and $p=0.3$ (the default
parameters used in \cite{Minai:ChaosNoiseCipher:Chaos1998} and
also used in the above sensitivity analysis), a return map
constructed from 9,000 samples of $K_t$ is shown in Fig.
\ref{figure:ReturnMap}a. It is clear that the two branches
correspond to the following two maps:
\begin{eqnarray}
K_{t+1} & = & \tanh(a_\mu K_t+A_\mu)-\tanh(b_\mu
K_t),\label{equation:ReturnMapA}\\
K_{t+1} & = & \tanh(a_\mu K_t+B_\mu)-\tanh(b_\mu
K_t),\label{equation:ReturnMapB}
\end{eqnarray}
respectively. In fact, with less samples it is still possible to
distinguish the two branches if they are not very close. Two
return maps constructed from 1,000 and 200 samples of $K_t$ are
shown in Figs. \ref{figure:ReturnMap}b and
\ref{figure:ReturnMap}c, respectively. Once the two branches are
distinguished, one can choose three points on each branch to try
to numerically solve the three secret parameters in the
corresponding equation.

\begin{figure}[!htbp]
\includegraphics{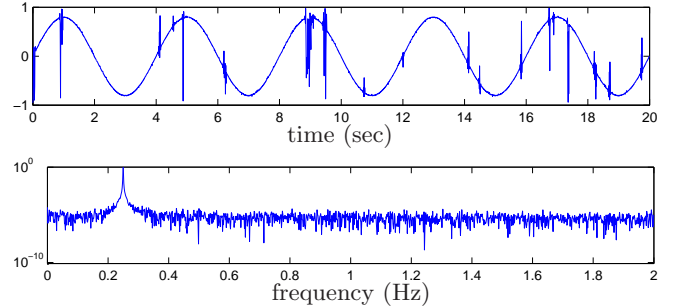}
\caption{The decrypted signal $\hat{m}_t$ and its relative power
spectrum with the estimated parameters, when $m_t=0.8\sin(2\pi
t/4)$, $p=0.3$ and $n=71$.}\label{figure:mt_mt2_sint_3_estimated}
\end{figure}

\begin{figure}[!htbp]
\includegraphics{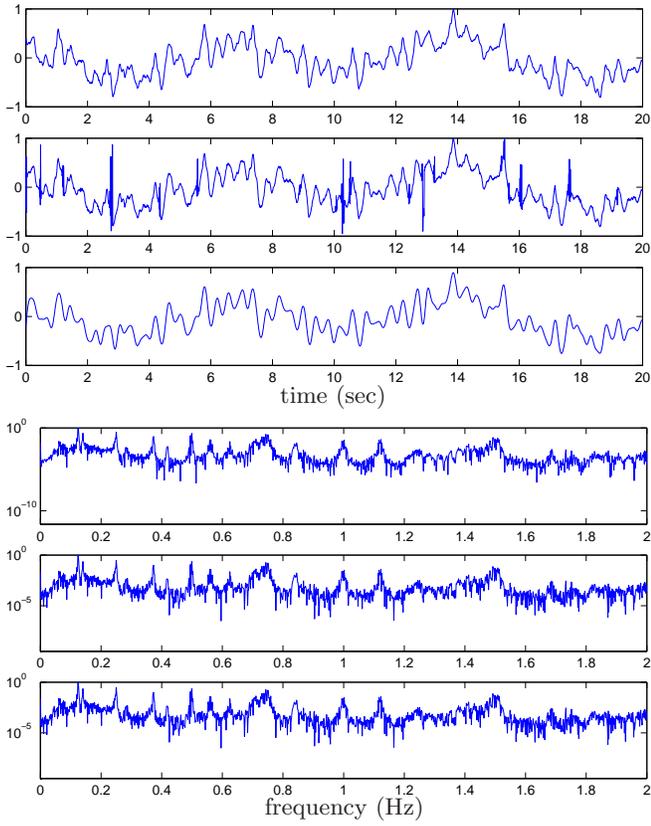}
\caption{The decrypted result with the estimated parameters, when
$m_t$ is a music file, $p=0.3$ and $n=71$ (from top to bottom: the
waveforms of $m_t$, $\hat{m}_t$ and the filtered $\hat{m}_t$, and
the relative power spectra of the three
signals).}\label{figure:mt_mt2_music_3_estimated}
\end{figure}

In \cite{Minai:ChaosNoiseCipher:Chaos1998}, it was claimed that
numerical solutions of Eqs. (\ref{equation:ReturnMapA}) and
(\ref{equation:ReturnMapB}) cannot work well due to the high
sensitivity of the decryption error to the parameters mismatch.
Unfortunately, as pointed out above in this paper, the decryption
error is actually not sufficiently sensitive to the parameter
mismatch. As a result, rough estimations of the secret parameters
can work well to generate a keystream $K_t^*\approx K_t$ for most
samples. One can directly get estimations of all the four secret
parameters with the following simple method discussed in
\cite{Minai:ChaosNoiseCipher:Chaos1998}:
\begin{itemize}
\item $A_\mu$ and $B_\mu$: the $y$-intercepts of the two branches
are $K_A^{0}=\tanh(A_\mu)$ and $K_B^{0}=\tanh(B_\mu)$,
respectively, so $\widetilde{A}_\mu=\tanh^{-1}(K_A^{0})$ and
$\widetilde{B}_\mu=\tanh^{-1}(K_B^{0})$;

\item $b_\mu$: since $a_\mu\geq 2 b_\mu$ \cite[Sec.
II]{Minai:ChaosNoiseCipher:Chaos1998}, the tails of both branches
will approach $f(x)=1-\tanh(b_\mu x)$ quickly as $x$ increases
(see Fig. \ref{figure:ReturnMap}a), thus $b_\mu$ can be
approximately derived from a point $(K_t,K_{t+1})$ lying in the
tail: $\widetilde{b}_\mu\approx\tanh^{-1}(1-K_{t+1})/K_t$;

\item $a_\mu$: once $b_\mu$ is approximately known, it is easy to
derive $a_\mu$ from a point $(K_t,K_{t+1})$ lying in the
$A$-branch, as follows:
$\widetilde{a}_\mu\approx(\tanh^{-1}(K_{t+1}+\tanh(\widetilde{b}_\mu
K_t))-\widetilde{A}_\mu)/K_t$, or a point $(K_t,K_{t+1})$ lying in
the $B$-branch:
$\widetilde{a}_\mu\approx(\tanh^{-1}(K_{t+1}+\tanh(\widetilde{b}_\mu
K_t))-\widetilde{B}_\mu)/K_t$.
\end{itemize}
Note that $\widetilde{b}_\mu$, $\widetilde{A}_\mu$ and
$\widetilde{B}_\mu$ depend only on the sampling points in the
return map, and that $\widetilde{a}_\mu$ depends on the estimation
of $b_\mu$ and $B_\mu$. This means that the error propagation only
occurs for $\widetilde{a}_\mu$. Since $a_\mu$ is relatively larger
than the other three parameters, it is less sensitive to the
estimation errors. Based on the only 200 samples shown in Fig.
\ref{figure:ReturnMap}c, the secret parameters are estimated as
follows:
\begin{itemize}
\item $K_A^{0}\approx 0.05\Rightarrow
\widetilde{A}_\mu=\tanh^{-1}(K_A^{0})\approx 0.05004172927849$,
and the relative estimation error is
$\delta_{A_\mu}=|\widetilde{A}_\mu-A_\mu|/A_\mu\approx 8.346\times
10^{-4}$;

\item $K_B^{0}\approx 0.7618\Rightarrow
\widetilde{B}_\mu=\tanh^{-1}(K_A^{0})\approx 1.00049031787725$,
and the relative estimation error is
$\delta_{B_\mu}=|\widetilde{B}_\mu-B_\mu|/B_\mu\approx 4.903\times
10^{-4}$;

\item the point $(K_3\approx 0.7624,K_4\approx 9.766\times
10^{-4})$ is used to derive
$\widetilde{b}_\mu\approx\tanh^{-1}(1-K_4)/K_3\approx
5.00000000000003$, and the relative estimation error is
$\delta_{b_\mu}=|\widetilde{b}_\mu-b_\mu|/b_\mu\approx 3\times
10^{-14}$;

\item the point $(K_{14}\approx 9.368\times 10^{-4},K_{15}\approx
0.0686)$ lying in $A$-branch is used to derive
$\widetilde{a}_\mu\approx(\tanh^{-1}(K_{15}+\tanh(\widetilde{b}_\mu
K_{14}))-\widetilde{A}_\mu)/K_{14}\approx 24.9554555$, and the
relative estimation error is
$\delta_{a_\mu}=|\widetilde{a}_\mu-a_\mu|/a_\mu\approx 1.782\times
10^{-3}$.
\end{itemize}
Recalling the weak sensitivity of the decryption error to
parameter mismatch when $p=0.3$, it can be expected that the above
estimation will achieve a rather good decryption performance.
Figure \ref{figure:mt_mt2_sint_3_estimated} gives the decryption
result when $m_t=0.8\sin(2\pi t/4)$. We have also tested the
decryption performance when $m_t$ is a music file (a PCM-encoded
16-bit wav file with the sampling frequency of 44kHz), where the
decrypted plain-signal $\hat{m}_t$ is further enhanced with a
low-pass filter. Figure \ref{figure:mt_mt2_music_3_estimated}
shows the decryption result, from which one can see that the
plain-music is almost perfectly reconstructed.

Note that only tens of samples may already be enough for
estimating the four parameters. When the number of samples is too
small, the two branches will not be clear. Fortunately, one can
still tell on which branch some points $(K_t,K_{t+1})$ lie if
$K_t\leq 0.1$ since in most cases the two branches separated are
sufficiently apart for $K_t\in[0,0.1]$.

Recalling the data shown in Table \ref{table:delta_max_p}, the
precision of the above estimation is good enough for all values of
$p$. Actually, even when the estimated parameters are not accurate
enough to get an acceptable decryption performance, the attacker
can still employ a more sophisticated algorithm to numerically
derive the parameters with a higher precision. In this case, the
estimated values can serve as good initial conditions for the
employed numerical solving algorithm. This implies that
\textit{the breaking of the secret parameters from a fragment of
$K_t$ is always possible, independent of the sensitivity of $K_t$
to the parameter mismatch}. In other words, \textit{the
key-generation process is always insecure against
known/chosen-plaintext attacks}. One can see that this result
agrees with the similar result obtained in the last subsection --
\textit{the security of the studied secure communication scheme is
ensured by the encryption function $\Psi^n(x,y)$ (not by the
chaos-based key-generation process)}. Once again, it discourages
the use of intermittent chaos in cryptography.

\subsection{The security of the enhanced key-generation process against known/chosen-plaintext attacks}

\begin{figure*}[!htbp]
\includegraphics{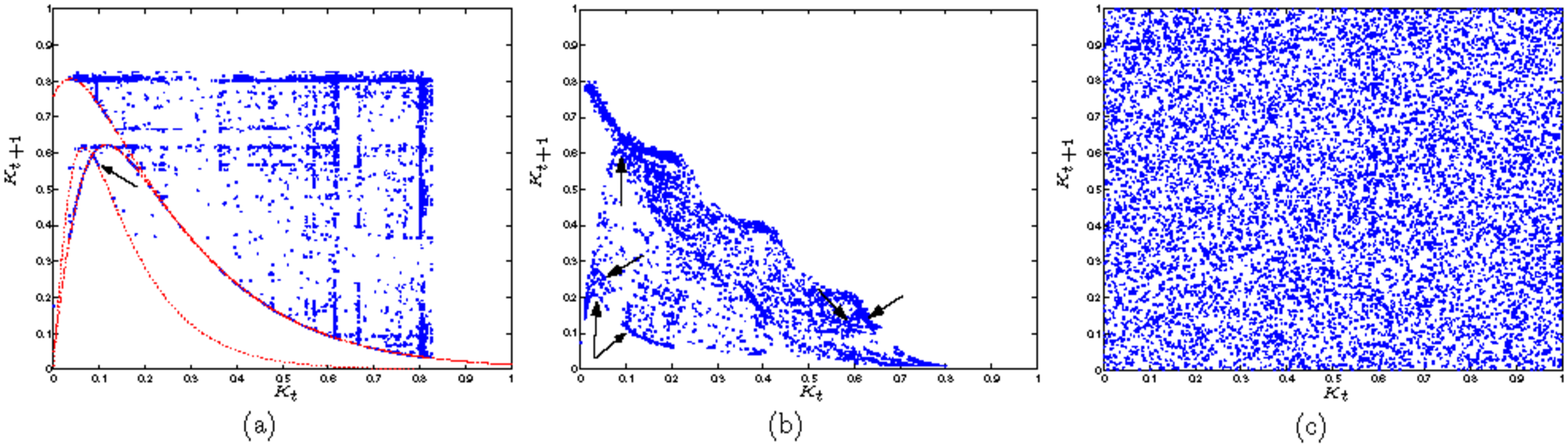}
\caption{The return maps constructed from 10,000 samples of $K_t$
in the enhanced key-generation process, when (a)
$K_t=\max_{k=1}^4(z_t^k)$, (b) $K_t=(z_t^1+z_t^2+z_t^3+z_t^4)/4$,
(c)
$K_t=T^{24}((z_t^1+z_t^2+z_t^3+z_t^4)/4,0.3)$.}\label{figure:ReturnMap_EnhancedScheme}
\end{figure*}

In this subsection, we consider the security of the enhanced
key-generation process proposed in
\cite{Minai:ChaosNoiseCipher:Chaos1998} against
known/chosen-plaintext attacks. Assume $m\;(m>1)$ different
dynamical systems are used to generate the keystream $K_t$:
$K_t=\max_{k=1}^m\left(z_t^k\right)$, where
\begin{equation}
z_{t+1}^k=\tanh\left(a_\mu^kz_t^k+u_\mu^k\right)-\tanh\left(b_\mu^kz_t^k\right).
\end{equation}
Apparently, in the enhanced key-generation process, the number of
secret parameters will increase to be $4m$ and the return map will
be distorted to some extent. Unfortunately, it is found that such
an enhancement is not as secure as expected against
return-map-based attacks. As an example, let $m=4$ and the
parameters be
$\{a_\mu^1,a_\mu^2,a_\mu^3,a_\mu^4\}=\{7.5,22.5,25,27.5\}$,
$\{b_\mu^1,b_\mu^2,b_\mu^3,b_\mu^4\}=\{2.5,4.5,5,5.5\}$,
$A_\mu^1=A_\mu^2=A_\mu^3=A_\mu^4=0.01$ and
$B_\mu^1=B_\mu^2=B_\mu^3=B_\mu^4=0.2$. The return map from 10,000
samples is plotted in Fig. \ref{figure:ReturnMap_EnhancedScheme}a.
Compared with Fig. \ref{figure:ReturnMap}a, although one cannot
find the shape of all the 8 branches corresponding to 8 different
equations, two of them can still be clearly seen, as shown by
dot-dashed lines (which are local edges of the return map). Two
exposed branches belong to the first dynamical sub-system:
$z_{t+1}=\tanh(a_\mu^1 z_t+A_\mu^1)-\tanh(b_\mu^1 z_t)$ and
$z_{t+1}=\tanh(a_\mu^1 z_t+B_\mu^1)-\tanh(b_\mu^1 z_t)$. Using the
method discussed above, one can easily get an estimation of the 4
secret parameters $\{a_\mu^1,b_\mu^1,A_\mu^1,B_\mu^1\}$. This
method can be further generalized to break even more parameters:
the several dots marked by the arrow actually expose another
branch, $z_{t+1}=\tanh(a_\mu^2 z_t+A_\mu^2)-\tanh(b_\mu^2 z_t)$,
which belongs to the second sub-system and is shown by a dotted
line in Fig. \ref{figure:ReturnMap_EnhancedScheme}a. Since no
point is available near the $y$-intercept of this branch, the
simple estimation method is disabled, but numerical algorithms can
still be used to get the values of $\{a_\mu^2,b_\mu^2,A_\mu^2\}$
by using only three different points. Now, about half of the
secret parameters are broken with the return-map-based
cryptanalysis. Although it seems very difficult to break other
secret parameters in a similar way, the breaking of partial
parameters still reduces the security of the enhanced
key-generation process quite significantly.

To frustrate the above partial attack, one may change
$K_t=\max_{k=1}^m\left(z_t^k\right)$ to other functions. In Fig.
\ref{figure:ReturnMap_EnhancedScheme}b, the return map
corresponding to the mean function
$K_t=(z_t^1+z_t^2+z_t^3+z_t^4)/4$ is shown. It can be seen that
the return map is further distorted and the edges become much more
ambiguous. However, there are still some visible curves (marked
with arrows) hidden in the noise-like return map. It is not clear
whether or not these visible curves can be used to break some
secret parameters. From a conservative point of view, the risk
always exists, so a good mixing function has to be found to remove
any visible information in the $K_t-K_{t+1}$ return map. We have
tested many simple functions and their combinations, and found
that it really is a difficult task. Considering the non-uniform
distribution of $K_t$ over the defining interval (see Fig. 3 of
\cite{Minai:NoiseSynchronization:PRE1998}), it is guessed that
only iterative chaotic maps with a good mixing property
\cite{Lasota:Stochastics-Chaos} can smooth the distribution of
$K_t$ and then effectively remove the visible curves. In Fig.
\ref{figure:ReturnMap_EnhancedScheme}c, the 24-fold skew-tent map
is used to generate the keystream:
$K_t=T^{24}((z_t^1+z_t^2+z_t^3+z_t^4)/4,0.3)$, where
\begin{equation}
T(x,p)=\begin{cases}
x/p, & 0\leq x\leq p,\\
(1-x)/(1-p), & p\leq x\leq 1.
\end{cases}
\end{equation}
One can see that the resulting return map becomes much more mixed.
Although in such a way the key-generation process can be
dramatically enhanced, \textit{the enhancement is caused by the
fully-chaotic tent map $T(x,p)$ with a good mixing feature, not by
the two intermittent chaotic systems themselves}. This, for the
third time, makes the use of intermittent chaos in cryptography
questionable.

\section{Conclusion}

This paper has carefully studied the security of a secure
communication scheme published in
\cite{Minai:ChaosNoiseCipher:Chaos1998}, which is based on
intermittent chaotic systems driven by a common random signal. It
is found that the key space of the studied scheme can be
drastically reduced, and that the decryption is insensitive to the
mismatch of the secret key, which means that the scheme can be
easily broken by inexpensive brute-force attacks. Furthermore, it
has been found that the core of this secure communication scheme
-- the key-generation process -- is not secure against
known/chosen-plaintext attacks: if an attacker can get access to a
fragment of the generated keystream, he can easily estimate the
secret key with sufficient accuracy and thus break the entire
cryptosystem. The security of an enhanced key-generation process
proposed in \cite{Minai:ChaosNoiseCipher:Chaos1998} has also been
discussed. At present, it is still not clear whether or not the
existence of periodic regimes in intermittent chaotic systems
always brings negative influence on the design of secure chaotic
cryptosystems. It is an open problem for future investigations.

\section*{Acknowledgements}

The authors would like to thank the anonymous reviewers for their
valuable comments and suggestions. This research was partially
supported by the Applied R\&D Centers of the City University of
Hong Kong under grants no. 9410011 and no. 9620004, and by the
Ministerio de Ciencia y Tecnolog\'{\i}a of Spain under research
grant SEG2004-02418.

\end{document}